\documentclass[submission,copyright,creativecommons]{eptcs}
\providecommand{\event}{AMMSE 2011} 

\usepackage{graphicx}
\usepackage{color}
\usepackage{url}
\usepackage{times}
\usepackage{AMMAlanguages}
\usepackage[all]{xy}

\newcommand{\code}[1]{{\small\texttt{#1}}}

\newcounter{marginalnote}
\newcommand{\mgcomment}[1]{}

\begin{document}

\pagestyle{headings} 

\title{Tracing Properties of UML and OCL Models\\with Maude}

\def\titlerunning{Tracing OCL with Maude} 

\author{
Francisco Dur\'an
\institute{ETSI Inform\'atica. Universidad de M\'alaga, Spain}
\and 
Martin Gogolla
\institute{University of Bremen, Germany}
\and 
Manuel Rold\'an
\institute{ETSI Inform\'atica. Universidad de M\'alaga, Spain}
}
\def\authorrunning{F. Dur\'an, M. Gogolla \& M. Rold\'an} 


\maketitle

\begin{abstract}

The starting point of this paper is a system described in form of a
UML class diagram where system states are characterized by OCL
invariants and system transitions are defined by OCL pre- and
postconditions. The aim of our approach is to assist the developer in
learning about the consequences of the described system states and
transitions and about the formal implications of the properties that
are explicitly given. We propose to draw conclusions about the stated
constraints by translating the UML and OCL model into the algebraic
specification language and system Maude, which is based on rewrite logic. We
will concentrate in this paper on employing Maude's capabilities for state
search. Maude's state search offers the possibility to describe a
start configuration of the system and then explore all configurations
reachable by rewriting. The search can be adjusted by formulating
requirements for the allowed states and the allowed transitions.

\end{abstract}


\section{Introduction}\label{sec:introduction}

In the last years, model-driven development has become the topic of
various research activities. The employment of models in all phases of
software development and for different purposes is said to offer a
higher level of abstraction than traditional code-centric
development. Models should concentrate on crucial properties of a
system to be developed or to be documented. Therefore, particular
attention must be paid to checking such properties in order to
guarantee software quality. For example, one is interested in
verifying whether the formulated model properties are consistent,
i.e., there are no contradictions between the properties, or one wants
to check whether some further property is implied by other fixed
properties. This paper will basically use UML and OCL to formulate
models and properties in form of OCL constraints.  OCL offers to
express properties regarding the system states by invariants and
properties concerning system transitions by pre- and
postconditions. Thus static and dynamic aspects can be handled.

Given a UML and OCL model of a system employing invariants and pre-
and postconditions as the central description means, the developer
frequently wants to know about the consequences on the described
system states and transitions and about the formal implications of the
properties that are explicitly given. For example, given some
invariant and some state, one wants to check whether another state,
where a further invariant does hold, can be reached by valid
transitions or not. Concerning operation pre- and postconditions, the
developer could wish to see all valid operation calls in a given state
which satisfy the preconditions. Another question is whether from a
given start state another explicitly given end state can be reached by
valid operations such that all intermediate states are respecting the
invariants. Questions like these and further ones will be handled in
our approach. For example, in our approach it is also possible to ask
for all sequences of messages and for all objects which have to be
created in order to reach a state satisfying specific OCL constraints.

On the technical level, we want to draw conclusions about the stated
constraints by translating the UML and OCL model into the algebraic
specification language and system Maude. Maude is based on rewrite logic,
is a well-established language, and offers sophisticated tools for the
analysis of specifications~\cite{CDELMMT:2007-book}. For example,
Maude allows the developer to check the satisfiability of LTL formulas 
using its model checker, or to prove system properties with the
inductive theorem prover ITP. However, we will concentrate in this
paper on employing Maude's capabilities for reachability analysis: 
we assume invariants are
represented by state predicates, operations by Maude rules and pre-
and postconditions by predicates as well. Basically, Maude's state
search functionality offers the possibility to describe a start configuration of the
system and then explore all configurations reachable from it by rewriting (or
a finite subset of all reachable configurations by imposing a maximal
depth on the graph of all possible configurations). The search can be
adjusted by formulating requirements for the allowed states and the
allowed transitions. To represent UML model in Maude and to evaluate OCL expressions on such models we use mOdCL~\cite{Roldan-Duran:2011}.

We assume the reader is familar with the Maude representation of
classes, objects and configurations as, for example, described
in~\cite{CDELMMT:2007-book}. We nevertheless provide a short description of the required Maude features in Section~\ref{sec:static}. The rest of this paper is structured as
follows. In Section~\ref{sec:example} we introduce our running example. Section~\ref{sec:static}
discusses the mOdCL representation of OCL types, user-defined UML classes,
and OCL constraints. Section~\ref{sec:behavior} concentrates on system dynamics by
introducing the general approach to handle operation calls. Section~\ref{sec:tracing}
shows how to exploit properties of the model by employing the Maude
state search. Section~\ref{sec:concl} finishes with concluding remarks and a brief
discussion of future work.

\section{Running Example}\label{sec:example}

Our running example describes a simple marriage world in which persons
can get married and can get divorced. In UML terms, we have a class
\code{Person} and two enumerations \code{Gender} (with literals
\code{female}, \code{male}) and \code{Civil\-Status} (with literals
\code{single}, \code{married}, \code{divorced}). In the \code{Person}
class we have attributes \code{civstat}, \code{gender}, \code{wife}
and \code{husband}. 

\begin{lstlisting}[style=AMMA, language=Maude, numbers=none]
class Person
attributes
  civstat:CivilStatus   gender:Gender         
  wife:Set(Person)      husband:Set(Person)
end
\end{lstlisting}

\begin{figure}[t]
 \begin{center}
 \includegraphics[width=0.8\textwidth]{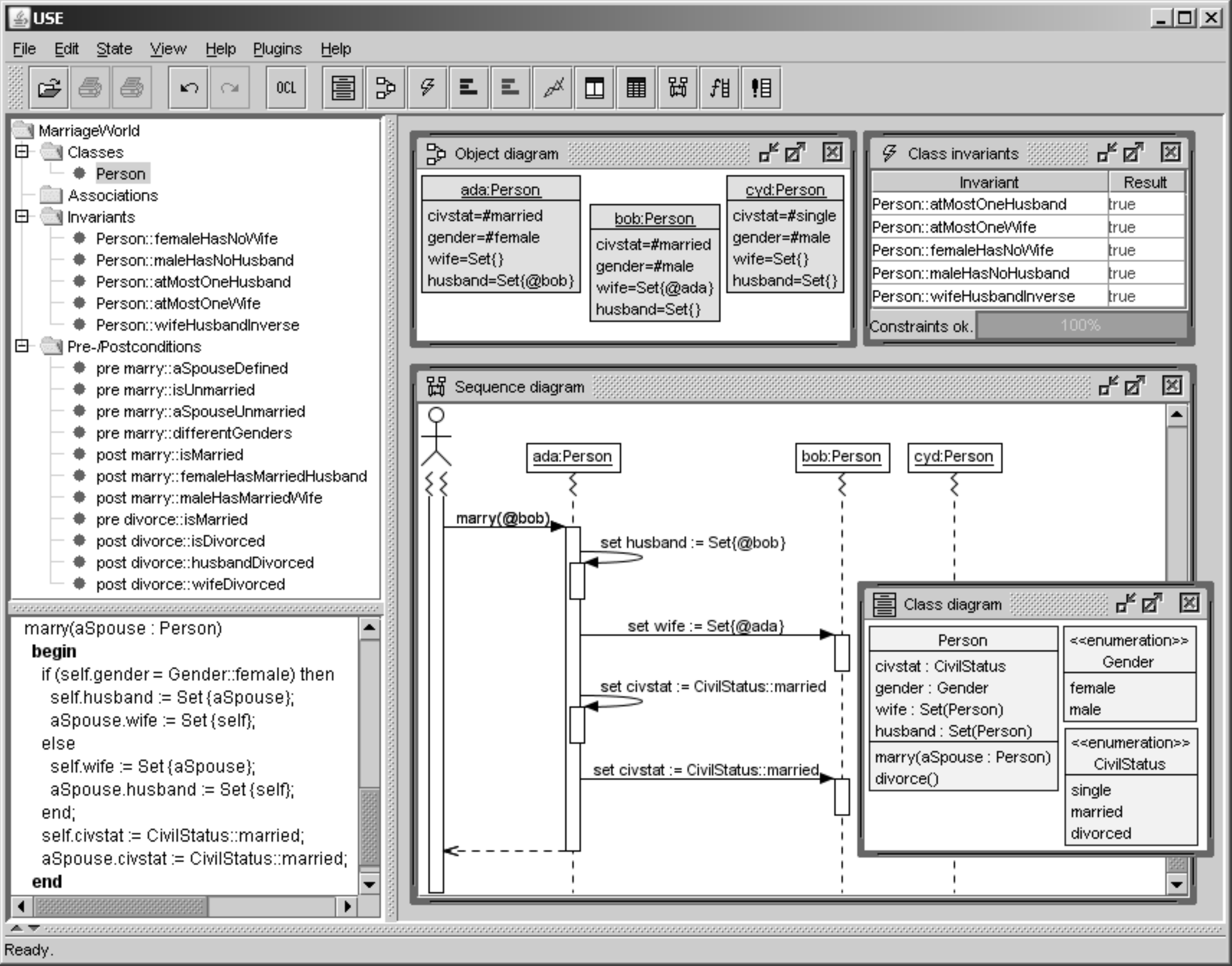}
 \caption{\label{USE-scenario}Snapshot of the USE tool for one of the example scenarios.}
 \end{center}
 \end{figure}

For the example, we have decided to present the spouses with
set-valued attributes, because we want to show how to use OCL expressions 
to avoid situations like polygamy and homosexual marriage, and because we wanted to avoid attributes being
undefined. If we would have single-valued attributes, e.g.,
\code{wife:Person}, then an unmarried male would be represented with
\code{wife} being equal to undefined, whereas in our model this is
represented as \code{wife} being equal to the empty set. In the
\code{Person} class there are two operations for marrying and
divorcing explained in more detail below.
We have developed and
checked our example with the UML and OCL tool
USE~\cite{Gogolla:2007:SCP}. See Figure~\ref{USE-scenario}.

As shown in the following, the states of the model are restricted by a
number of OCL invariants and the transitions in the model, the
operation calls, are required to obey certain operation contracts in
form of OCL pre- and postconditions.

A person should not be married to a person of the same gender
(\code{female\-Has\-No\-Wife} and \code{male\-Has\-No\-Husband}), a
person can have at most one spouse (\code{atMost\-One\-Husband} and
\code{atMost\-One\-Wife}), and marriage is a symmetric association
(\code{wife\-Husband\-Inverse}). We can express these invariants for
the \code{Person} class in OCL:

\begin{lstlisting}[style=AMMA, language= OCL20, numbers=none]
 inv femaleHasNoWife: gender=female implies wife->isEmpty()
 inv maleHasNoHusband: gender=male implies husband->isEmpty()
 inv atMostOneHusband: husband->size()<=1
 inv atMostOneWife: wife->size()<=1
 inv wifeHusbandInverse:
       (wife->notEmpty() implies wife.husband->includes(self)) and
       (husband->notEmpty() implies husband.wife->includes(self))
\end{lstlisting}

In order to guarantee consistency of the attribute values, one could additionally state the invariant \code{(civstat=married) = (wife->notEmpty() or husband->notEmpty())}.

Class \code{Person} has a \code{marry(aSpouse:Person)} operation. If
(a)~the argument \code{aSpouse} of the operation is
defined,\footnote{In OCL, there is an exception element representing
undefined; it is possible to pass this undefined element as an
argument to a \code{marry} call.} (b)~both the person receiving
the message and the argument \code{aSpouse} in the message are not
married, and (c)~both persons have different gender, then the operation
results in a state where both persons are married. The operation can
be characterized as follows in OCL:\footnote{Given the first part of the consequent \code{husband=Set\{aSpouse\}} in the \code{femaleHasMarriedHusband} constraint, one
could simplify its second part to \code{aSpouse.civstat=married} assuming
that the formal operation parameter \code{aSpouse} cannot be changed by the
operation. And similarly for the \code{maleHasMarriedWife} constraint. However, we prefer in our example the explicit formulation.
}

\begin{lstlisting}[style=AMMA, language= OCL20, numbers=none]
 pre  aSpouseDefined: aSpouse.isDefined
 pre  isUnmarried: civstat<>married
 pre  aSpouseUnmarried: aSpouse.civstat<>married
 pre  differentGenders: gender<>aSpouse.gender
 post isMarried: civstat=married
 post femaleHasMarriedHusband: gender=female implies
      husband=Set{aSpouse} and husband.civstat->forAll(cs|cs=married)
 post maleHasMarriedWife: gender=male implies
      wife=Set{aSpouse} and wife.civstat->forAll(cs|cs=married)
\end{lstlisting}

Two married persons can get divorced using the \code{divorce()}
operation.  This operation can be specified in OCL as follows:

\begin{lstlisting}[style=AMMA, language= OCL20, numbers=none]
 pre  isMarried: civstat=married
 post isDivorced: civstat=divorced
 post husbandDivorced: gender=female implies
      husband->isEmpty() and husband@pre.civstat->forAll(cs|cs=divorced)
 post wifeDivorced: gender=male implies
      wife->isEmpty() and wife@pre.civstat->forAll(cs|cs=divorced)
\end{lstlisting}

\section{The mOdCL Representation of Class Diagrams}\label{sec:static}

The representation of UML models used in mOdCL is described in
detail in~\cite{Roldan-Duran:2011}. This representation is inspired by the
representation of object-oriented modules in Maude, as those used in
Maude-based metamodeling frameworks such as
MOMENT~\cite{Boronat-Meseguer:08}, Maudeling~\cite{RRDV:07-jot}, and
e-Motions~\cite{Rivera-Duran-Vallecillo:2009}. In these approaches,
class diagrams are represented as Maude object-oriented modules, and
system states as configurations of objects. The representation used in
mOdCL has its own particularities, mainly in the representation of
attributes, links and methods, and of course in the support of OCL. In
comparison to other Maude-like approach to UML and
OCL~\cite{DBLP:conf/ecmdafa/BoronatOGRC06,DBLP:conf/amast/ClavelE06},
mOdCL is the only Maude implementation providing support
for pre- and postconditions, and the only one supporting the dynamic validation of OCL constraints.

We briefly describe in the following sections the representation of
class diagrams and OCL constraints used in mOdCL illustrating it with
our running example.

\subsection{Object-Oriented Programming in Maude}
\label{sec:object-based}

Maude~\cite{CDELMMQ:2002,CDELMMT:2007-book} is a high-level language and a high-performance system that supports membership equational logic and rewriting logic specification and programming of systems.

Membership equational logic~\cite{Meseguer:1998}, a Horn logic whose atomic sentences are equalities $t=t'$ and \emph{membership assertions} of the form $t : S$, stating that a term $t$ has sort $S$. 
Such a logic extends order-sorted equational logic, and supports sorts, subsort relations, subsort polymorphic overloading of operators, and the definition of partial functions with equationally defined domains.

Rewriting logic~\cite{Meseguer:1992-tcs} is a logic of change
that can naturally deal with state and with highly nondeterministic
concurrent computations.
In rewriting logic,  the state space of a distributed system is specified as an
algebraic data type in terms of an equational specification
$(\Sigma,E)$, where $\Sigma$ is a signature of sorts (types) and
operations, and $E$ is a set of equational axioms. The dynamics of a
system in rewriting logic is then specified by rewrite \emph{rules} of the form
$t \rightarrow t'$, where $t$ and $t'$ are $\Sigma$-terms.
This rewriting happens modulo the equations $E$, describing in fact
local transitions $[t]_E\rightarrow[t']_E$.
These rules describe the local, concurrent transitions possible in the
system, i.e. when a part of the system state fits the pattern $t$
(modulo the equations $E$)
then it can change to a new local state fitting pattern $t'$.

In Maude, concurrent object-oriented systems can be naturally modeled in Maude as multisets of objects and messages,
loosely coupled by some suitable communication mechanism. The basic sorts needed to describe an object system are: \code{Object}, \code{Msg} (messages), and \code{Configuration}. 

Given a class $C$ with attributes $a_1 \texttt{:} S_1 \texttt{,}
\; ... \texttt{,} \; a_n \texttt{:} S_n$, where 
$a_i$ are attribute identifiers and $S_i$ are the sorts of the corresponding
attributes, objects of such a class $C$ are then record-like structures of sort \code{Object} of the form
$\texttt{<} \; O \; \texttt{:} \; C \; \texttt{|} \; a_1 \texttt{:} v_1
\texttt{,} \; ... \texttt{,} \; a_n \texttt{:} v_n \; \texttt{>}$, where $O$ is
the (unique) identifier of the object, and $v_i$ are the current values of its attributes.

Objects can interact in a number of different ways, including message passing.
Messages are declared in Maude as operations of sort \code{Msg}. 

To represent objects, additional sorts are available: \code{Oid} (object identifiers), \code{Cid} (class identifiers),
\code{Attribute} (named elements of objects' states), and \code{AttributeSet} (comma-separated multisets of
attributes). Thus, for example, for a class \code{Account} with a single attribute \code{balance} of sort \code{Int}, 
and messages \code{withdraw} and \code{transfer} we have the following declarations:

\begin{lstlisting}[style=AMMA, language=MaudeModule, numbers=none]
 sort Account .
 subsort Account < Cid .
 op Account : -> Account [ctor] .
 op balance :_ : Int -> Attribute [ctor] .
  
 op withdraw : Oid Int -> Msg [ctor] .
 op transfer : Oid Oid Int -> Msg [ctor] .
\end{lstlisting}
  
Thus, given the predefined operator 

\begin{lstlisting}[style=AMMA, language= MaudeModule, numbers=none]
 op <_:_|_> : Oid Cid AttributeSet -> Object [ctor] .
\end{lstlisting}
  
\noindent we can represent an account object with identifier \code{a} with balance $5$ as a term 

\begin{lstlisting}[style=AMMA, language=Maude, numbers=none]
 < a : Account | balance : 5 >
\end{lstlisting}
  
\noindent and a message to \code{a} to withdraw $3$ as

\begin{lstlisting}[style=AMMA, language=Maude, numbers=none]
 withdraw(a, 3)
\end{lstlisting}

In a concurrent object-oriented system, the concurrent state, which is called a
\emph{configuration}, has the structure of a multiset made up of objects and
messages that evolves by concurrent rewriting using rules that describe the
effects of the communication events of objects and messages. The predefined
sort \code{Configuration} represents configurations of Maude objects and
messages, with \code{none} as empty configuration and the empty syntax
operator {\small\verb~__~} as union of configurations.

\begin{lstlisting}[style=AMMA, language= MaudeModule, numbers=none]
 sort Configuration .
 subsorts Object Message < Configuration .
 op none : -> Configuration [ctor] .
 op __ : Configuration Configuration -> Configuration
     [ctor assoc comm id: none] .
\end{lstlisting}

Thus, such rewrite rules define transitions between configurations, and their
general form is:

{\small \ttfamily
\begin{tabbing}
\ \ \ \ cr\=l [\textit{r}] :\\
      \> < \(O_1\) : \(C_1\) | \(\textit{atts}_1\) > \(...\) < \(O_n\) : \(C_n\) | \(\textit{atts}_n\) >\\
      \> \(M_1\) \(...\) \(M_m\)\\
      \> => \= < \(O_{i_1}\) : \(C'_{i_1}\) | \(\textit{atts}'_{i_1}\) > \(...\) < \(O_{i_k}\) : \(C'_{i_k}\) | \(\textit{atts}'_{i_k}\) >\\
      \>    \> < \(Q_1\) : \(C''_1\) | \(\textit{atts}''_1\) > \(...\) < \(Q_p\) : \(C''_p\) | \(\textit{atts}''_p\) >\\
      \>    \> \(M'_1\) \(...\) \(M'_q\)\\
      \> if \textit{Cond} .
\end{tabbing}
\rmfamily }

\noindent where $r$ is the rule label, $M_1... M_m$ and $M'_1... M'_q$ are
messages, $O_1... O_n$ and $Q_1... Q_p$ are object identifiers, $C_1... C_n$,
$C'_{i_1}... C'_{i_k}$ and $C''_1... C''_p$ are classes, $i_1... i_k$ is a
subset of $1...n$, and $Cond$ is a Boolean condition (the rule's \emph{guard}).
The result of applying such a rule is that: $(a)$ messages $M_1... M_m$
disappear, i.e., they are consumed; $(b)$ the state, and possibly the classes
of objects $O_{i_1}... O_{i_k}$ may change; $(c)$ all the other objects $O_j$
vanish; $(d)$ new objects $Q_1... Q_p$ are created; and $(e)$ new messages
$M'_1... M'_q$ are created, i.e., they are sent. Rule labels and guards are
optional.
When several objects or messages appear in the left-hand side of a rule, they
need to synchronize in order for such a rule to be fired. 

Class inheritance is directly supported by
Maude's order-sorted type structure. A subclass declaration \code{C < C'},
indicating that \code{C} is a subclass of \code{C'}, is a particular case
of a subsort declaration \code{C < C'}, by which all attributes, messages,
and rules of the superclasses, as well as the newly defined attributes,
messages and rules of the subclass characterize its structure and behavior.
Multiple inheritance is also supported in
Maude~\cite{CDELMMT:2007-book}.

\subsection{OCL Basic Types}
\label{sec:basic-types}

Basic predefined OCL types are directly mapped on predefined Maude
sorts. Thus, the OCL types \code{Boolean}, \code{Integer}, \code{Real}
and \code{String} are mapped into the Maude predefined sorts
\code{Bool}, \code{Int}, \code{Float}, and \code{String},
respectively.  User-defined classes are mapped on the Maude predefined
sort \code{Oid} (of object identifiers). OCL predefined collection
types are mapped into sorts \code{Set}, \code{Bag}, \code{OrderedSet},
and \code{Sequence}. All these sorts are subsorts of \code{OclType}.



In OCL, the \code{Enum} sort is intended to be used to define
enumeration types. Thus, types \code{Gender} and \code{CivilStatus}
are declared as enumerations by declaring them as subsorts of
\code{Enum}. The values of the types are then declared as constants of
the respective types. E.g., the declarations for \code{CivilStatus} are:

\begin{lstlisting}[style=AMMA, language= MaudeModule, numbers=none]
 sort CivilStatus .                          
 subsort CivilStatus < Enum .
 ops single married divorced : -> CivilStatus [ctor] . 
\end{lstlisting}

Classes are represented following the standard representation of
classes in Maude (see Section~\ref{sec:static} and \cite{CDELMMT:2007-book}). The \code{Attribute} sort
provides the syntax for attributes and role ends in the objects, given
by its name, of sort \code{AttributeName}, and its type, of sort
\code{OclType}.

\begin{lstlisting}[style=AMMA, language=MaudeModule, numbers=none]
 op _:_ : AttributeName OclType -> Attribute [ctor] .
\end{lstlisting}

\noindent The name of a given class \textit{C} is represented as a
constant \code{\it C} of the Maude sort \code{Cid}, and its attributes
and role ends as constants of the mOdCL sort \code{AttributeName}.
Thus, for the \code{Person} class we add declarations:

\begin{lstlisting}[style=AMMA, language=MaudeModule, numbers=none]
 sort Person .
 subsort Person < Cid .
 op Person : -> Person [ctor] .  
 ops civstat gender husband wife : -> AttributeName [ctor] .
\end{lstlisting}


Role ends with multiplicity 1 are represented as attributes of sort
\code{Oid}, and role ends with multiplicity \code{*} as attributes of
sort \code{Set} (for \code{Oid} sets).


An operation
{\small\texttt{\textit{op}(\(\textit{arg}_1\):\(\textit{type}_1\),\(\ldots\),\(\textit{arg}_n\):\(\textit{type}_n\)):
\textit{type}}} is represented as a constant
\texttt{\small\textit{op}}, of sort \texttt{\small OpName} (of
operation names), and constants {\small\texttt{\(\textit{arg}_1\)},
$\ldots$, \texttt{\(\textit{arg}_n\)}}, of sort {\small\texttt{Arg}} (of
arguments).  Thus, operations \code{marry(aSpouse:Person)} and
\code{divorce()} require the following declarations:

\begin{lstlisting}[style=AMMA, language= MaudeModule, numbers=none]
 ops marry divorce : -> OpName [ctor] .
 op aSpouse : -> Arg [ctor] .
\end{lstlisting}

Once we have defined the structural part, we can create object configurations. 
%
%
%
%
%
Since we will be using these configurations in several commands in the
following sections, we define constants \code{abc-single} and
\code{ab-married-c-single}
representing, respectively, the object configurations in which persons
\code{ada}, \code{bob}, and \code{cyd} are single, and \code{ada} and
\code{bob} married and \code{cyd} single, respectively.
      
\begin{lstlisting}[style=AMMA, language= MaudeModule, numbers=none]
 ops ada bob cyd : -> Oid .
 ops abc-single ab-married-c-single : -> Configuration .
 eq abc-single 
   = < ada : Person | gender : female, civstat : single,  wife : Set{},    husband : Set{} > 
     < bob : Person | gender : male,   civstat : single,  wife : Set{},    husband : Set{} > 
     < cyd : Person | gender : male,   civstat : single,  wife : Set{},    husband : Set{} > .
 eq ab-married-c-single 
   = < ada : Person | gender : female, civstat : married, wife : Set{},   husband : Set{bob} > 
     < bob : Person | gender : male,   civstat : married, wife : Set{ada}, husband : Set{} > 
     < cyd : Person | gender : male,   civstat : single,  wife : Set{},    husband : Set{} > .
\end{lstlisting}


\subsection{OCL Constraints Representation}
\label{sec:mOdCL-constraints}

In mOdCL, OCL expressions are represented as terms of sort
\code{OclExp}. We can represent the OCL expressions from
Section~\ref{sec:example} as corresponding mOdCL terms of sort
\code{OclExp}. The mOdCL terms mirror quite closely the OCL
expressions given in Section~\ref{sec:example}. We give here a few of
them for illustration purposes:

\begin{lstlisting}[style=AMMA, language= MaudeModule, numbers=none]
 op femaleHasNoWife : -> OCL-Exp .
 eq femaleHasNoWife = (context Person inv gender = female implies wife -> isEmpty()) .
\end{lstlisting}


\begin{lstlisting}[style=AMMA, language= MaudeModule, numbers=none]
 op atMostOneHusband : -> OCL-Exp .
 eq atMostOneHusband = (context Person inv husband -> size() <= 1) .
\end{lstlisting}


\begin{lstlisting}[style=AMMA, language= MaudeModule, numbers=none]
 op wifeHusbandInverse : -> OCL-Exp .
 eq wifeHusbandInverse 
   = context Person inv 
       (wife -> notEmpty() implies wife . husband -> includes(self)) and
       (husband -> notEmpty() implies  husband . wife -> includes(self)) .
\end{lstlisting}

\begin{lstlisting}[style=AMMA, language= MaudeModule, numbers=none]
 op aSpouseDefined : -> OCL-Exp .
 eq aSpouseDefined = (aSpouse . isDefined()) .
\end{lstlisting}





\begin{lstlisting}[style=AMMA, language= MaudeModule, numbers=none]
 op femaleHasMarriedHusband : -> OCL-Exp .
 eq femaleHasMarriedHusband 
   = (gender = female implies 
        husband = Set{aSpouse} and husband . civstat -> forAll(cs | cs = married)) .
\end{lstlisting}



\begin{lstlisting}[style=AMMA, language= MaudeModule, numbers=none]
 op husbandDivorced : -> OCL-Exp .
 eq husbandDivorced 
   = gender = female implies
       (husband -> isEmpty() and husband @pre . civstat -> forAll(cs | cs = divorced)) .
\end{lstlisting}


\noindent 

Notice the use of constants (like \code{female\-Has\-No\-Wife}) to
identify these expressions. Invariants and pre- and postconditions are
then specified by operators \code{inv}, \code{pre} and \code{post} in
order to build the \code{OclExp} to be checked when the validation
process requires to inspect invariants, pre- or postconditions of a
given method, respectively.

\begin{lstlisting}[style=AMMA, language= MaudeModule, numbers=none]
 op inv : -> OclExp .
 ops pre post : OpName -> OclExp .
\end{lstlisting}

The global invariant to be satisfied is given as an operator
\code{inv} defined as the conjunction of all invariant expressions.
Given the above definitions, the \code{inv} constant can be defined
as:

\begin{lstlisting}[style=AMMA, language= MaudeModule, numbers=none]
 eq inv = femaleHasNoWife and maleHasNoHusband and atMostOneHusband and 
          atMostOneWife and wifeHusbandInverse .
\end{lstlisting}

The \code{pre} and \code{post} operators must be defined for each
method. These operators need the associated operation as an
argument. We only show the corresponding declarations for the \code{marry} operation; it works analogously for \code{divorce}.

\begin{lstlisting}[style=AMMA, language= MaudeModule, numbers=none]
 eq pre(marry) = (aSpouseDefined and isUnmarried and aSpouseUnmarried and differentGenders) .
 eq post(marry) = (isMarried and femaleHasMarriedHusband and maleHasMarriedWife) .
\end{lstlisting}

With this structural description of the system, we can evaluate any
OCL expression on object configurations by using the mOdCL's
\code{eval} operator. We can for instance evaluate the \code{inv} OCL
expression defined above on the \code{abc-single} object
configuration introduced in Section~\ref{sec:static}. 

\begin{lstlisting}[style=AMMA, language=Maude, numbers=none]
 Maude> red eval(inv, abc-single) .
 result Bool: true
\end{lstlisting}


\section{System Dynamics}
\label{sec:behavior}

\subsection{Representation of Operation Calls}

Once the static part of the system is completely determined, we can
specify its behavior.  We could do that in different ways, but our
goal is to be able to generate execution traces leading to states
satisfying or violating given OCL constraints.  However, making sure
that all invariants and pre- and postconditions are satisfied in the
right places, is not an easy task.  To discharge the user of the
burden, mOdCL provides facilities to perform the appropriate checks in
the appropriate places just by following very simple rules.
Basically, the preconditions of an operation must be checked before
its execution starts, and when it is completed, its postconditions and
the invariants must be satisfied.  To be able to automatically perform
the checks before and after the execution of each operation, mOdCL
expects that methods are invoked with a call message of the form

\begin{lstlisting}[style=AMMA, language=Maude, numbers=none]
 call(<method-name>, <addressee>, <argument-list>)
\end{lstlisting}
and upon their completion they send a return message of the form
\begin{lstlisting}[style=AMMA, language=Maude, numbers=none]
 return(<return-value>)
\end{lstlisting}
The infrastructure of mOdCL will intercept these messages and will 
take the following measures. 

The processing of a \code{call} operator results in the execution of a
method, for which a context object, representing the execution context
with the appropriate information for the running method, is
generated. This object, of class \code{Context}, is of the form
\begin{lstlisting}[style=AMMA, language=Maude, numbers=none]
 < Ctx : Context | op : M, self : Id, args : Vars >
\end{lstlisting}
where \code{Ctx} is the identifier of the object, \code{M} is the name
of the active method, \code{Id} is the identifier of the current
object, and \code{Vars} is a set of (variable name,value) pairs
corresponding to the arguments of the method invocation and local
variables.  The modeler can make use of this \code{Context} object to
get or manipulate the value of an argument variable or a local
variable.

A return message \code{return(<return-value>)} will be replaced 
by a resume message of the form
\begin{lstlisting}[style=AMMA, language=Maude, numbers=none]
 resume(<return-value>)
\end{lstlisting}

To manage the chaining of method invocations~--- an invoked method can
invoke another one (or recursively itself)~--- the validator uses an
execution stack in which the necessary information is
stored.\footnote{mOdCL currently assumes a single thread of execution. 
A multithreaded execution would also be possible just by adding support for multiple stacks.}  mOdCL configurations are terms of sort
\code{Con\-fi\-gu\-ra\-tion+}, which is declared as a sort with a
single constructor

\begin{lstlisting}[style=AMMA, language=MaudeModule, numbers=none]
 op {_} : Configuration -> Configuration+ .
\end{lstlisting}

where one of the objects in the wrapped configuration represents the
execution stack, which is represented by a \code{stack} operator.

\subsection{Operations \code{marry} and \code{divorce}}

With these restrictions, we can specify the behavior of our operations
as follows.  Since the husband and wife attributes are sets of object
identifiers, the \code{marry} operation just adds a new husband or a
new wife to the corresponding set, depending on the gender of the
receiver of the message.

\begin{lstlisting}[style=AMMA, language=MaudeModule, numbers=none]
 vars Ctx Self Dude : Oid .   vars St1 St2 : CivilStatus .
 vars Gd1 Gd2 : Gender .      vars Hb1 Hb2 Wf1 Wf2 : Set .
\end{lstlisting}

\begin{lstlisting}[style=AMMA, language=MaudeRule, numbers=none]
 rl [MARRY] :
  < Ctx : Context | op : marry, self : Self, args : arg(aSpouse, Dude) >
  marry-token
  < Self : Person | civstat : St1, gender : Gd1, husband : Hb1, wife : Wf1 > 
  < Dude : Person | civstat : St2, gender : Gd2, husband : Hb2, wife : Wf2 >
  => if Gd1 == female 
     then < Self : Person | civstat : married, gender : Gd1, husband : Set{Dude}, wife : Wf1 > 
          < Dude : Person | civstat : married, gender : Gd2, husband : Hb2, wife : Set{Self} > 
     else < Self : Person | civstat : married, gender : Gd1, husband : Hb1, wife : Set{Dude} > 
          < Dude : Person | civstat : married, gender : Gd2, husband : Set{Self}, wife : Wf2 >
     fi 
     < Ctx : Context | op : marry, self : Self, args : arg(aSpouse, Dude) >
     return(0) .
\end{lstlisting}

In the case of the \code{divorce} operation, the set of the
corresponding \code{husband} or \code{wife} attribute is just left
empty.  We use two rules to distinguish cases, depending on the
gender, but show only the rule for females.

\begin{lstlisting}[style=AMMA, language=MaudeModule, numbers=none]
 vars Ctx Self Dude : Oid .        vars St1 St2 : CivilStatus .        var  Wf : Set .
 vars Gd1 Gd2 : Gender .           vars Atts1 Atts2 : AttributeSet .
\end{lstlisting}

\begin{lstlisting}[style=AMMA, language=MaudeRule, numbers=none]
 rl [DIVORCE] :
   < Ctx : Context | op : divorce, self : Self, args : empty >
   divorce-token
   < Self : Person | civstat : St1, gender : female, husband : Set{Dude}, Atts1 >
   < Dude : Person | civstat : St2, wife : Wf, Atts2 >
   => < Self : Person | civstat : divorced, gender : female, husband : Set{}, Atts1 >
      < Dude : Person | civstat : divorced, wife : Set{}, Atts2 >
      < Ctx : Context | op : divorce, self : Self, args : empty >
      return(0) .
\end{lstlisting}


These methods are very simple, and they are both
specified so that their execution will consists in the application of
one rule.  The \code{marry-token} and \code{divorce-token} operators
are used to prevent the execution of the rules corresponding to the
operation out of the scope of a call to such an operation.  More
intricate methods may require several rules to be executed
sequentially, in which case more than a simple token will be required.

Methods are invoked with \code{call} operators. However, for
convenience, and to use a notation more friendly, we will use the
following operators.

\begin{lstlisting}[style=AMMA, language=MaudeModule, numbers=none]
 op _. marry(_) : Oid Oid -> Msg [msg] .
 op _. divorce() : Oid -> Msg [msg] .
\end{lstlisting}

These methods will be used to produce the sending of a corresponding
\code{call} message.

\begin{lstlisting}[style=AMMA, language=MaudeRule, numbers=none]
 rl [MARRY-INVOCATION] :
   Self . marry(Arg)
   stack(nil)
   => call(marry, Self, arg(aSpouse, Arg))   
      stack(nil)   
      marry-token .

 rl [DIVORCE-INVOCATION] :
   Self . divorce()   
   stack(nil)
   => call(divorce, Self, empty)   
      stack(nil)   
      divorce-token .  
\end{lstlisting}

Notice the presence of the \code{stack} operator, with \code{nil}
contents in both rules, to guarantee that no new message is sent until
the execution stack is empty, that is, the processing of all previous
messages has been completed.

Given this behavior we can simulate the system by using the
\code{rewrite} Maude command,\footnote{The \code{rewrite} Maude
command (abbreviated \code{rew}) causes the specified term to be
rewritten using the rules, equations, and membership axioms in the
given module.} or we can explore all possible executions searching for
states satisfying a specific property.

\begin{lstlisting}[style=AMMA, language= MaudeRule, numbers=none]
 Maude> rew { abc-single   
              cyd . marry(ada)   
              stack(nil) } .
 result Configuration+: 
   { stack(nil) 
     resume(marry, 0) 
     < ada : Person | civstat : married, gender : female, husband : Set{cyd}, wife : Set{} > 
     < bob : Person | civstat : single, gender : male, husband : Set{}, wife : Set{} > 
     < cyd : Person | civstat : married, gender : male, husband : Set{}, wife : Set{ada} > }
\end{lstlisting}

What happens if we try to do something that does not respect the
defined constraints?  For instance, if we send a \code{marry} message
to an already married person, the precondition of the \code{marry}
operation should fail. mOdCL will detect situations like this and will
send corresponding error messages.
   
\begin{lstlisting}[style=AMMA, language=MaudeRule, numbers=none]
 Maude> rew { ab-married-c-single 
              cyd . marry(ada) 
              stack(nil) } .
 result [Configuration+]: 
   { marry-token 
     < ada : Person | civstat : married, gender : female, husband : Set{bob}, wife : Set{} > 
     < bob : Person | civstat : married, gender : male, husband : Set{}, wife : Set{ada} > 
     < cyd : Person | civstat : single, gender : male, husband : Set{}, wife : Set{} > 
     error("Precondition violation", 
           marry, 
           aSpouse . isDefined() and civstat <> married and
           gender <> aSpouse . gender and aSpouse . civstat <> married) }
\end{lstlisting}

The error message provides a string explaining the failure, and the
OCL expression that has failed.  In the case of pre- and
postconditions, the name of the operation of the condition is also
indicated.

\section{Tracing Properties}\label{sec:tracing}

Notice that the configuration rewritten is a set, and so, if several
messages were in it, no order in which they are to be consumed is
assumed.  If we were interested in a sequence of messages with a
specific order, we could very easily specify it.  But this
associative-commutative configuration allows us to explore the
different traces in which certain messages may be consumed.  For
instance, the following search command looks up all possible traces
leading to states in which there are two persons married using
messages \code{cyd . marry(ada)}, \code{ada . marry(bob)}, and
\code{bob . divorce()}.

\begin{lstlisting}[style=AMMA, language=Maude, numbers=none]
 Maude> search { abc-single 
                 (cyd . marry(ada))
                 (ada . marry(bob)) 
                 (bob . divorce()) 
                 stack(nil) }
           =>* { stack(nil) 
                 Conf:Configuration } 
           such that
               eval(Person . allInstances -> exists(P |
                      Person . allInstances -> exists(Q |
                        P . wife -> includes(Q) and Q . husband -> includes(P))),
                    Conf:Configuration) .
 Solution 1 (state 4)
 Conf:Configuration 
   --> bob . divorce() 
       resume(marry, 0) 
       (cyd . marry(ada)) 
       < ada : Person | civstat : married, gender : female, husband : Set{bob}, wife : Set{} > 
       < bob : Person | civstat : married, gender : male, husband : Set{}, wife : Set{ada} > 
       < cyd : Person | civstat : single, gender : male, husband : Set{}, wife : Set{} >
 Solution 2 (state 5)
 Conf:Configuration 
   --> bob . divorce() 
       resume(marry, 0) 
       (ada . marry(bob)) 
       < ada : Person | civstat : married, gender : female, husband : Set{cyd}, wife : Set{} > 
       < bob : Person | civstat : single, gender : male, husband : Set{}, wife : Set{} > 
       < cyd : Person | civstat : married, gender : male, husband : Set{}, wife : Set{ada} >
 Solution 3 (state 12)
 Conf:Configuration 
   --> resume(marry, 0) 
       resume(marry, 0) 
       resume(divorce, 0) 
       < ada : Person | civstat : married, gender : female, husband : Set{cyd}, wife : Set{} > 
       < bob : Person | civstat : divorced, gender : male, husband : Set{}, wife : Set{} > 
       < cyd : Person | civstat : married, gender : male, husband : Set{}, wife : Set{ada} >
 No more solutions.
\end{lstlisting}

The Maude search command gives us all possible reachable states
satisfying the specified OCL expression.  More interestingly, we have
access to the paths followed for each of these solutions.  By using
the \code{show path} and \code{show path labels} commands, we can get
all the information on the path followed up to any of the states in
the search space, and specifically to the ones given as solutions of
the search command~--- notice the state number given on the right of
the solution number.

\begin{lstlisting}[style=AMMA, language=Maude, numbers=none]
 Maude> show path labels 12 . 
 MARRY-INVOCATION
 MARRY
 DIVORCE-INVOCATION
 DIVORCE
 MARRY-INVOCATION
 MARRY
\end{lstlisting}

\subsection{Spontaneous Generation of Messages}

This is very powerful already, but we can do more.  In the above
search we had to give what messages we wanted to use to generate all
possible paths.  But that does not guarantee that we get all possible
ways to get a specific state. We now present a possible alternative in
which new messages are spontaneously generated.  As a result of the
execution we are interested in getting the sequence of messages
leading to the searched states.  To obtain it, we add a
\code{messages} operator that keeps the list of messages generated.

\begin{lstlisting}[style=AMMA, language=MaudeModule, numbers=none]
 sort MsgList .
 subsort Msg < MsgList . 
 op nilMsgList : -> MsgList .
 op _;_ : MsgList MsgList -> MsgList [assoc id: nilMsgList] .

 op messages : MsgList -> Msg [ctor] .
\end{lstlisting}

Consider the following rules, by which, if there exist two persons, a
new \code{marry} message for the marriage of these two persons is
generated:

\begin{lstlisting}[style=AMMA, language= MaudeModule, numbers=none]
 var  ML : MsgList .                vars Dude1 Dude2 : Oid .
 vars Atts1 Atts2 : AttributeSet .  var  Arg : OclType .
  
 rl [NEW-MARRY-MESSAGE] : 
   < Dude1 : Person | Atts1 >
   < Dude2 : Person | Atts2 >
   messages(ML)
   => < Dude1 : Person | Atts1 >
      < Dude2 : Person | Atts2 >
      hold(messages(ML ; Dude1 . marry(Dude2)))
      Dude1 . marry(Dude2) .
\end{lstlisting}

In addition to sending a new \code{marry} message, the rule adds the
generated message to the list of messages. And, since we do not want
to generate a new message until the previous one was consumed, we
block the \code{messages} operator with a \code{hold} operator.

\begin{lstlisting}[style=AMMA, language= MaudeModule, numbers=none]
 op hold : Msg -> Msg [ctor] .
\end{lstlisting}

Once the operation has finished, we can unblock the message list.

\begin{lstlisting}[style=AMMA, language= MaudeRule, numbers=none]
 var  ML : MsgList .      var  Arg : OclType .
  
 rl [RETURN-MARRY] : 
   resume(marry, Arg)   
   hold(messages(ML))
   => messages(ML) .
\end{lstlisting}     

Similar rules are added to spontaneously generate new
\code{divorce} messages.

\mgcomment{In the case of the \code{divorce} operation,
to consider the two alternative genders, we write two rules for the
generation, and one more for the \code{return} message:

\begin{lstlisting}[style=AMMA, language=Maude, numbers=none]
 vars Dude1 Dude2 : Oid .      var  Atts : AttributeSet .
 var  Arg : OclType .          var  ML : MsgList .
 vars MS : List .
  
 rl [NEW-DIVORCE-MESSAGE-1] : 
   < Dude1 : Person | husband : Set{Dude2 ; MS}, Atts >
   messages(ML)
   => < Dude1 : Person | husband : Set{Dude2 ; MS}, Atts >
      Dude1 . divorce() 
      hold(messages(ML ; Dude1 . divorce())) .
 rl [NEW-DIVORCE-MESSAGE-2] : 
   < Dude1 : Person | wife : Set{Dude2 ; MS}, Atts >
   messages(ML)
   => < Dude1 : Person | wife : Set{Dude2 ; MS}, Atts >
      Dude1 . divorce() 
      hold(messages(ML ; Dude1 . divorce())) .

 rl [RETURN-DIVORCE] :
   resume(divorce, Arg)
   hold(messages(ML))
   => messages(ML) .
\end{lstlisting}     
}


Notice the use of the messages operator to collect the sequence of
messages generated.  When a new message is generated it is added to
the list of messages in the messages operator.  With it we will know
the messages used to reach a given state.  We could instead use the show path
command to get the sequence of rules, and from it obtain the sequence
of messages, but notice that the search command looks for states, and
only keeps the shortest path to each reached state.

We can get, e.g., the states reached when searching for married
couples.  Of course, the number of reachable states satisfying this
condition is infinite, and we must either limit the number of
solutions to look for or the maximum depth of the search.  We can
search for the first 10,000 states as follows:


\begin{lstlisting}[style=AMMA, language=Maude, numbers=none]
 Maude> search [10000] 
          { stack(nil) 
            abc-single 
            messages(nilMsgList) } 
          =>* { Conf:Configuration messages(ML:MsgList) } 
          such that
              eval(Person . allInstances -> exists(P |
                     Person . allInstances -> exists(Q | 
                       P . wife -> includes(Q) and Q . husband -> includes(P))),
                   Conf:Configuration) .
 Solution 1 (state 17)
 Conf:Configuration 
   --> stack(nil) 
       < ada : Person | civstat : married, gender : female, husband : Set{bob}, wife : Set{} > 
       < bob : Person | civstat : married, gender : male, husband : Set{}, wife : Set{ada} > 
       < cyd : Person | civstat : single, gender : male, husband : Set{}, wife : Set{} >
 ML:MsgList --> ada . marry(bob)

 ...

 Solution 10000 (state 131696)
 Conf:Configuration 
   --> stack(nil) 
       < ada : Person | civstat : married, gender : female, husband : Set{cyd}, wife : Set{} > 
       < bob : Person | civstat : divorced, gender : male, husband : Set{}, wife : Set{} > 
       < cyd : Person | civstat : married, gender : male, husband : Set{}, wife : Set{ada} >
 ML:MsgList --> ada . marry(cyd) ; 
                cyd . divorce() ; 
                bob . marry(ada) ; 
                bob . divorce() ; 
                cyd . marry(ada) ; 
                cyd . divorce() ; 
                ada . marry(cyd) ;
                ada . divorce() ; 
                cyd . marry(ada)
\end{lstlisting}

We may also look for more interesting states. 
For example, we may search for states in which the
precondition of an operation fails.  We must again limit the search.

\begin{lstlisting}[style=AMMA, language=Maude, numbers=none]
 Maude> search [20,20] 
          { stack(nil)
            abc-single 
            messages(nilMsgList) }
          =>* { error(Str:String, ON:OpName, Exp:OCL-Exp) 
                Conf:[Configuration] 
                hold(messages(ML:MsgList)) } .              
 Solution 1 (state 10)
 Conf:[Configuration] 
   --> marry-token 
       < ada : Person | civstat : single, gender : female, husband : Set{mt}, wife : Set{mt} > 
       < bob : Person | civstat : single, gender : male, husband : Set{mt}, wife : Set{mt} > 
       < cyd : Person | civstat : single, gender : male, husband : Set{mt}, wife : Set{mt} >
 ML:MsgList --> bob . marry(cyd)
 Str:String --> "Precondition violation"
 ON:OpName --> marry
 Exp:OCL-Exp 
   --> aSpouse . isDefined() and civstat <> married and 
       gender <> aSpouse . gender and aSpouse . civstat <> married
...
 Solution 20 (state 74)
 Conf:[Configuration] 
   --> marry-token 
     < ada : Person | civstat : married, gender : female, husband : Set{bob}, wife : Set{mt} >  
     < bob : Person | civstat : married, gender : male, husband : Set{mt}, wife : Set{ada} > 
     < cyd : Person | civstat : single, gender : male, husband : Set{mt}, wife : Set{mt} >
 ML:MsgList --> (bob . marry(ada)) ; cyd . marry(bob)
 Str:String --> "Precondition violation"
 ON:OpName --> marry
 Exp:OCL-Exp --> aSpouse . isDefined() and civstat <> married and 
                 gender <> aSpouse . gender and aSpouse . civstat <> married
\end{lstlisting}
% ...

However, we can show how we cannot reach states in which invariants are not satisfied. 

\begin{lstlisting}[style=AMMA, language=Maude, numbers=none]
 Maude> search [20,20] 
          { stack(nil)
            abc-single 
            messages(nilMsgList) }
          =>* { error(Str:String, Exp:OCL-Exp) 
                Conf:[Configuration] 
                hold(messages(ML:MsgList)) } . 
 No solution.
\end{lstlisting}

Of course, since we have limited the search, the absence of solutions
does not imply that such states do not exist.  But notice that the
reachable state space is infinite because of the messages operator.
Without it, the state space is in fact small for our example.  We can
verify properties on our example by using an equational abstraction
\cite{Meseguer-Palomino-Marti-Oliet:2008} consisting in just removing
the messages in the messages list.

\begin{lstlisting}[style=AMMA, language=MaudeModule, numbers=none]
 var ML : MsgList .

 ceq messages(ML) = messages(nilMsgList) if ML =/= nilMsgList .    
\end{lstlisting}

We can now, for example, prove that within our Maude description no
reachable state fails the invariant.

\begin{lstlisting}[style=AMMA, language=Maude, numbers=none]
 Maude> search { stack(nil) 
                 abc-single 
                 messages(nilMsgList) } 
           =>* { error(Str:String, Exp:OCL-Exp) 
                 Conf:[Configuration] 
                 hold(messages(ML:MsgList)) } . 
 No solution.
\end{lstlisting}

Or more interestingly, that after you marry you will never be single again.

\begin{lstlisting}[style=AMMA, language=Maude, numbers=none]
 Maude> search { stack(nil)
                 ab-married-c-single 
                 messages(nilMsgList)] 
           =>* { Conf:[Configuration] 
                 messages(ML) } 
           such that 
               eval(ada . civstat = single or bob . civstat = single, Conf) .
 No solution.
\end{lstlisting}

\subsection{Spontaneous Generation of Objects}

But why limiting ourselves to playing with \code{ada}, \code{bob}, and
\code{cyd}?  In the same way we can spontaneously generate new
messages to consider all possible sequences of messages, we can also
think on the possibility of spontaneously generating new objects.  In
fact, what we need is just a \code{new} or \code{create} message that
gives place to the object.  We declare a \code{new} message with the
identifier of the object, its class, and its gender to generate \code{Person} objects.

\begin{lstlisting}[style=AMMA, language=MaudeModule, numbers=none]
 op new : Oid Cid Gender -> Msg [ctor] .
\end{lstlisting}

As in the generation of messages, to generate a new object we need to
initialize its attributes.  In our current example, they are generated
single, and the only values we need to worry about are their
identifier and their gender.  Genders can be assigned
nondeterministically.  We give two rules to generate objects of each
of the possible genders. Regarding identifiers, we use a
\code{counter} operator to both indexing object identifiers $id(1)$,
$id(2)$, ... and limiting the number of generated objects.

\begin{lstlisting}[style=AMMA, language=MaudeModule, numbers=none]
 op id : Nat -> Oid [ctor] .
 op counter : Nat -> Msg [ctor] .
\end{lstlisting}
  

\begin{lstlisting}[style=AMMA, language=MaudeModule, numbers=none]
 var ML : MsgList .     var N : Nat .     var G : Gender .     var O : Oid .
  
 ops P Q : -> Vid .
  
 rl [NEW-MALE] : 
   counter(s N) 
   messages(ML)
   => counter(N) 
      new(id(N), Person, male) 
      hold(messages(ML ; new(id(N), Person, male))) .
 rl [NEW-FEMALE] : 
   counter(s N) 
   messages(ML)
   => counter(N) 
      new(id(N), Person, female) 
      hold(messages(ML ; new(id(N), Person, female))) .
 rl [PERSON-GENERATION] :
   new(O, Person, G)
   hold(messages(ML))
   => < O : Person | civstat : single, gender : G, husband : Set{}, wife : Set{} >
      messages(ML) .
\end{lstlisting}


We can now look for traces reaching states with married couples as
follows:

\begin{lstlisting}[style=AMMA, language=Maude, numbers=none]
 Maude> search [100,30] { stack(nil)
                          messages(nilMsgList) 
                          counter(3) } 
           =>* { Conf:Configuration messages(ML:MsgList) } 
           such that 
               eval(Person . allInstances -> exists(P | 
                      Person . allInstances -> exists(Q | 
                        P . wife -> includes(Q) and Q . husband -> includes(P))),
                    Conf) .
 Solution 1 (state 121)
 Conf:Configuration 
   --> 
   stack(nil) 
   counter(1) 
   < id(1) : Person | civstat : married, gender : female, husband : Set{id(2)}, wife : Set{} > 
   < id(2) : Person | civstat : married, gender : male, husband : Set{}, wife : Set{id(1)} >
 ML:MsgList --> new(id(2), Person, male) ; 
                new(id(1), Person, female) ;
                id(1) . marry(id(2))
 Solution 2 (state 122)
 Conf:Configuration
   -->
   stack(nil) 
   counter(1) 
   < id(1) : Person | civstat : married, gender : female, husband : Set{id(2)}, wife : Set{} > 
   < id(2) : Person | civstat : married, gender : male, husband : Set{}, wife : Set{id(1)} >
 ML:MsgList --> new(id(2), Person, male) ; 
                new(id(1), Person, female) ;
                id(2) . marry(id(1))
 ...
 Solution 100 (state 2420)
 Conf:Configuration 
   --> 
   stack(nil) 
   counter(0) 
   < id(1) : Person | civstat : married, gender : female, husband : Set{id(0)}, wife : Set{} > 
   < id(2) : Person | civstat : single, gender : male, husband : Set{}, wife : Set{} > 
   < id(0) : Person | civstat : married, gender : male, husband : Set{mt}, wife : Set{id(1)} >
 ML:MsgList --> new(id(2), Person, male) ; 
                new(id(1), Person, female) ;
                new(id(0), Person, male) ; 
                id(1) . marry(id(0)) ;
                id(0) . divorce() ; 
                id(0) . marry(id(1))
\end{lstlisting}

We can look for persons married to themselves.

\begin{lstlisting}[style=AMMA, language=Maude, numbers=none]
 Maude> search [20,20] { stack(nil)   messages(nilMsgList)   counter(3) } 
                 =>* { Conf:Configuration 
                       messages(ML:MsgList) } 
          such that eval(Person . allInstances -> exists(P | 
                           P . wife -> includes(P) or P . husband -> includes(P)),
                        Conf:Configuration) .
 No solution.
\end{lstlisting}

\section{Conclusions}\label{sec:concl}

We have made a proposal to draw consequences about stated constraints
in a UML and OCL model by translating the model into Maude.  Maude
allows the developer to check for system properties in various ways.
We have employed the Maude state search and were able to show that
particular states can (resp.~cannot) be reached under given assumptions,
i.e., by limiting the search to a finite subset of possible
configurations. We have generated scenarios, i.e., state sequences and
transitions, where the transition sequences were automatically
constructed in order to achieve a path from a start state to an end
state.

We believe Maude with its sophisticated tools offers more and even
stronger possibilities than the one we have employed. The Maude ITP
theorem prover could be used to achieve general propositions about
invariants. The Maude temporal logic checker could be employed for
verifying and finding complex states and transition
constellations. Fine grained information about constraint failure
would be desirable for mOdCL. Larger case studies must give feedback
about the practicability of our approach. All results obtained on the
Maude level must be translated back into UML and OCL, for example, in
terms of object and sequence diagrams.


\bibliographystyle{eptcs}
\bibliography{duran}

\end{document}